\documentclass[11pt]{article}
\RequirePackage{amsthm,amsmath,amssymb, bm}
\usepackage{mathbbol}
\usepackage{graphicx,psfrag,epsf}
\usepackage{enumerate}
\usepackage[round]{natbib}
\usepackage{url} % not crucial - just used below for the URL 
\usepackage{subfigure}
\usepackage[colorlinks,citecolor=blue,urlcolor=blue]{hyperref}
\usepackage{mathtools}
\usepackage{mathrsfs} 
\usepackage{booktabs}
\usepackage{multirow}

\newcommand{\E}{{\mathrm{E}}}
\newcommand{\pr}{{\mathrm{P}}}

\newcommand{\rom}[1]{\uppercase\expandafter{\romannumeral #1\relax}}

\newcommand{\ind}{\perp\!\!\!\!\perp}

\newtheorem{assumption}{Assumption}

\title{\bf Nonparametric Bounds in Causal Mediation Analysis in the Presence of Unmeasured Confounding and Imperfect Compliance}
\author{Wei Liang \ and \ Changbao Wu}
\date{} % Clear date explicitly

\begin{document}
\def\spacingset#1{\renewcommand{\baselinestretch}
{#1}\small\normalsize} \spacingset{1}

\maketitle % Only use \maketitle once
\renewcommand{\thefootnote}{}
\footnotetext{Emails: \texttt{w62liang@uwaterloo.ca} \ and \ \texttt{cbwu@uwaterloo.ca}}

\begin{abstract}
The average causal mediation effect (ACME) and the natural direct effect (NDE) are two parameters of primary interest in causal mediation analysis. However, the two causal parameters are not identifiable from randomized experimental data in the presence of outcome-mediator confounding and treatment-assignment noncompliance. Sj\"{o}lander (2009) addressed the partial identification issue and derived nonparametric bounds of the NDE in randomized controlled trials under a set of monotonicity assumptions based on the Balke-Pearl algorithm. These bounds provide partial information on the parameters and can be used for sensitivity analysis. In this paper, we extend Sj\"{o}lander's bounds on the NDE as well as the ACME to randomized controlled trials in the presence of noncompliance when the treatment assignment serves as an instrumental variable. Nonparametric sharp bounds for the local causal parameters defined on the subpopulation of treatment-assignment compliers are also established. We demonstrate the practical usefulness of the proposed upper and lower bounds through an application to the randomized experimental dataset on Job Search Intervention Study.
\end{abstract}

\spacingset{1.45}

%\newpage
\section{Introduction}
\label{sec:intro}

Causal mediation analysis plays a crucial role in many scientific fields, including social sciences, health and psychological researches, financial and economical studies, and beyond. Unlike traditional causal inference which focuses primarily on effects of a treatment on specific outcomes over the target population or subpopulations, causal mediation analysis digs into the underlying causal mechanisms by exploring how the treatment influences outcomes through mediators. This approach shifts the focus from merely identifying treatment effects to understanding the causal pathways that explain how the treatment produces its effects. 

%We focus on causal mediation analysis in randomized controlled trials. Specifically, we consider the setting described in \citet{angrist1996identification} and \citet{balke1997bounds} where treatment assignments are randomized but imperfect compliance to the assignments introduces confounding bias. The treatment assignment serves as an instrumental variable in this setting assuming the exclusion-restriction assumption. Such scenario is very common in randomized experiments \citep{cuzick1997adjusting,imbens1997bayesian}, where the practitioners typically perform an ``intent-to-treat'' analysis to evaluate the effect of the treatment assignment. However, the naive ``intent-to-treat'' analysis can result in misleading conclusions if the treatment effectiveness rather than the assignment effectiveness is of primary interest \citep{pearl2009causality}.  
%The noncompliance bias can also be viewed as a result of treatment-mediator and treatment-outcome confounding.

There are two central parameters in causal mediation analysis \citep{pearl2001direct,imai2010identification}: the average causal mediation effect (ACME), which operates through the mediator, and the natural direct effect (NDE), which captures the effect through other pathways. Standard randomization on treatment assignments, however, does not guarantee identification of the ACME and the NDE unless strong ignorability assumptions on the mediator are imposed, which require either full observations of pre-treatment mediator-outcome confounding or that the values of the mediator do not alter the direct effect of the treatment on the outcome \citep{pearl2001direct,robins2003semantics,imai2010general,imai2010identification,imai2013experimental}. Such requirements are stringent, untestable, and often fragile in practice. 

\citet{sjolander2009bounds} addressed the issue of identification by deriving nonparametric bounds on the NDE in randomized controlled trials under a set of monotonicity assumptions when the outcomes, mediators, and treatments are all binary, using the Balke-Pearl algorithm \citep{balke1994counterfactual,balke1995probabilistic}. These bounds do not rely on the ignorability assumptions and can be narrow and informative in certain settings. Even when the bounds are wide, they provide partial information on the causal parameters of interest and can be used for sensitivity analysis \citep{richardson2015nonparametric}.

%Extensive research has been conducted on partial identification in causal inference. For example, \citet{manski1990nonparametric} showed nonparametric bounds of the average treatment effect in two-arm randomized trials with imperfect compliance, and its sharp bounds when the outcomes are binary are given by \citet{balke1997bounds}. \citet{cheng2006bounds} extended their results to average treatment effects within principal strata in three-arm randomized experiments. For causal mediation analysis in randomized trials, \citet{sjolander2009bounds} derived the bounds on the NDE. The bounds on other causal parameters such as the controlled direct and indirect effects also have been studied in either randomized experiments or observational studies \citep{cai2008bounds, vanderweele2011controlled}. For a comprehensive review of the partial identification results on average treatment effects using instrumental variables, readers can refer to \citet{swanson2018partial}.

In this paper, we extend Sj\"{o}lander's bounds to randomized controlled trials in the presence of noncompliance, which refers to individuals who do not conform to treatment assignments. Noncompliance is common in randomized experiments due to ethical or logistical issues. For instance, the experimenter is usually not allowed to force a candidate to receive an assigned treatment. In the presence of noncompliance, extra outcome-treatment confounding bias is introduced when treatment effectiveness rather than assignment effectiveness is of primary interest \citep{pearl2009causality}, and to the best of our knowledge, no prior research has established bounds on the ACME and the NDE for this scenario. We consider the setting described in \citet{angrist1996identification} and \citet{balke1997bounds} where the treatment assignment serves as an instrumental variable under the exclusion-restriction assumption. We derive nonparametric bounds on the ACME and the NDE and also their local counterparts which are defined on the subpopulation of treatment assignment compliers. Furthermore, we consider a set of monotonicity assumptions which lead to tighter bounds of the causal parameters when the imposed assumptions hold.

We organize the rest of the paper as follows. In Section \ref{sec:setup}, we formulate the problem of interest and explain the assumptions underlying our model. Sections \ref{sec:bounds-pop} and \ref{sec:bounds-local} derive nonparametric bounds on the causal parameters of interest, with Section \ref{sec:bounds-pop} presenting the bounds on population causal parameters and Section \ref{sec:bounds-local} focusing on the bounds on local causal parameters. In Section \ref{sec:real-data}, we apply the bounds to a randomized experimental dataset to illustrate their practical usefulness. Section \ref{sec:remarks} concludes the paper with additional remarks on potential refinements to the proposed bounds in the presence of covariates. Technical details and proofs as well as some additional results are provided in the Appendix.

\section{Problem Setup}
\label{sec:setup}

\subsection{Data Generating Mechanism}
\label{subsec:setup-dgp}

Consider causal mediation analysis with experimental data. The individuals in the study are randomly assigned to two treatment groups, indicated by a random variable $Z\in\{0,1\}$. With possible assignment noncompliance, the actual treatment status is given by $A\in\{0,1\}$. We define $A(1)$ and $A(0)$, based on the Rubin causal model (RCM) \citep{rubin1974estimating}, as the two potential outcomes of $A$, i.e., the values of $A$ would have been under respectively $Z=1$ and $Z=0$. Then, the pair $(A(1),A(0))$ has four potential values, with $(1,0)$ corresponding to (treatment-assignment) compliers and $(0,1)$ corresponding to defiers. Noncompliance refers to individuals who are not compliers. 

Let $Y$ denote the outcome and $M$ a mediator of interest, both measured after the treatment. We further assume that $Y$ and $M$ are also dichotomous, taking value either $0$ or $1$. It is often convenient to let $1$ represent a favorable result. The temporal sequence of the variables is $Z\rightarrow A\rightarrow M\rightarrow Y$. We use the example in \citet{sjolander2009bounds} to illustrate our setup, where it is of interest to study the effect of smoking on cardiovascular disease (CVD) through hyperlipidemia. We further suppose that smokers are recruited and the intervention of Nicotine Replacement Therapy (NRT) is randomly assigned to help the smokers quit smoking ($Z=1$). However, the intervened smokers might not adhere to the instruction and persist in smoking during the study ($A(1)=0$), while smokers in the other group may quit smoking without NRT ($A(0)=1$). At the end of the study, the hyperlipidemia and CVD are measured, where $M=1$ and $Y=1$ respectively represent no hyperlipidemia and no CVD. 

The temporal relationship allows us to define potential outcomes for $Y$ and $M$. For $z,a,m\in\{0,1\}$, we assume $Y(z,a,m), Y(a,m)$ are the potential outcomes of $Y$ under respectively $(Z,A,M)=(z,a,m)$ and $(A,M)=(a,m)$, and $M(z,a)$ and $M(a)$ are the potential outcomes of $M$ under respectively $(Z,A)=(z,a)$ and $A=a$. We consider the following three assumptions throughout this paper.

\begin{assumption}[Exclusion restriction]\label{assp:exc-restr}
For any   $(z,a,m)\in\{0,1\}^3$, we have $Y(z,a,m)=Y(a,m)$ and $M(z,a)=M(a)$.
\end{assumption}

\begin{assumption}[Relevance]\label{assp:relev}
   $\E[A(1)-A(0)]\neq 0$.
\end{assumption}

\begin{assumption}[Randomized assignment]\label{assp:rand}
   $$Z\ind \{Y(1,1),Y(1,0),Y(0,1),Y(0,0),M(1),M(0),A(1),A(0)\}\,,$$
where ``$\ind$'' means ``is independent of''.   
\end{assumption}

\begin{figure}[t!]
    \centering
    \includegraphics[width=0.32\textwidth]{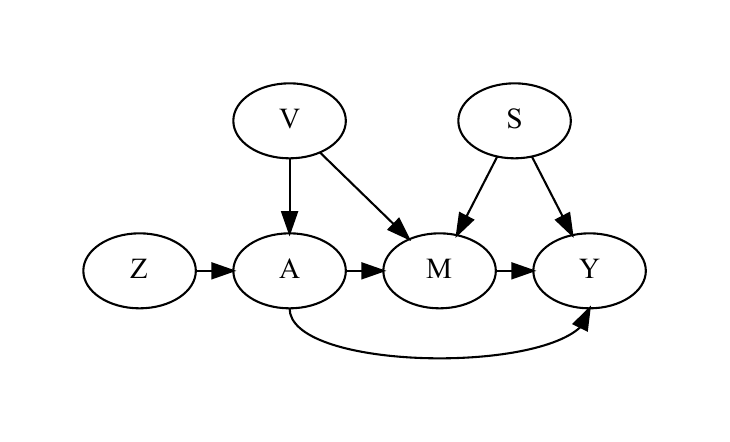}
    \includegraphics[width=0.32\textwidth]{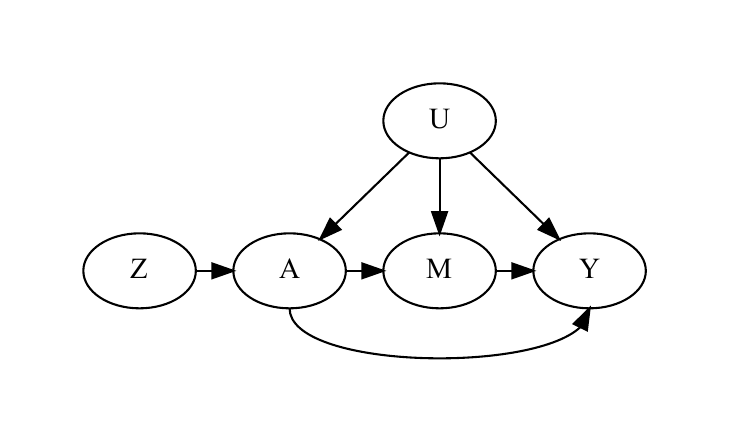}
    \includegraphics[width=0.32\textwidth]{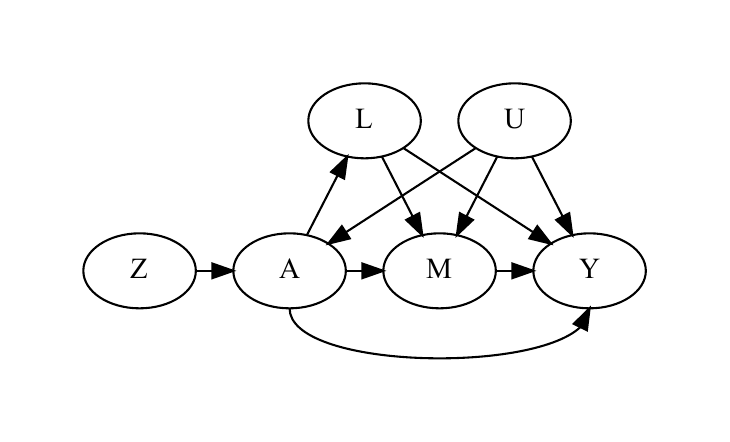}
    \caption{Feasible causal diagrams of the data generating mechanism under Assumptions \ref{assp:exc-restr}--\ref{assp:rand}.}
    \label{fig:CausalDiagram}
\end{figure}

Under the above three assumptions, $Z$ serves as an instrumental variable (IV) for the causal effects of $A$ on $M$ and $Y$ in the sense of \citet{pearl2009causality} under the structural causal models. In the sense of \citet{angrist1996identification}, Assumption \ref{assp:monotone-A} as given in Section \ref{subsec:setup-addass} is additionally required to ensure that $Z$ is an IV such that there are no defiers. To make the presentation clear, we refer to the model under Assumptions \ref{assp:exc-restr}--\ref{assp:rand} as ``the IV setting'', and the model under Assumptions \ref{assp:exc-restr}--\ref{assp:monotone-A} as ``the IV setting without defiers''. Assumption \ref{assp:relev} states that the assignment $Z$ has an overall impact on the treatment $A$, which is very weak and often desirable in practice. Assumption \ref{assp:rand} is a natural result of the randomization on $Z$ at the design stage. The most controversial part is Assumption \ref{assp:exc-restr}, which requires that the causal path from $Z$ to $M$ must be through $A$, and the one from $Z$ to $Y$ must be through $A$ or $M$. This assumption is untestable, assessing which requires background knowledge and arguments based on practical situations. In the smoking example, this assumption is plausible if we believe that the treatment of NRT does not affect the lipid levels of a person provided that the person actually smoke or not. \citet{angrist1996identification} includes comprehensive discussions on this assumption. 
%When there are alternative IVs for the causal effect, the IV estimates under different IV specifications can be computed and compared to evaluate the validity of the IVs \citep{angrist1995two}. The Hansen J overidentification test may be used for testing if multiple estimates are equivalent \citep{hansen1982large}, despite rejection of the alternative hypothesis based on which does not mean that the IVs are valid.
Assumptions \ref{assp:exc-restr}--\ref{assp:rand} are the basis of our setting for randomized controlled trials with noncompliance and admit a wide range of causal diagrams for the data, with three instances given in Figure \ref{fig:CausalDiagram}. We allow for the presence of unmeasured treatment-mediator confounding, treatment-outcome confounding, mediator-outcome confounding, as well as treatment-mediator-outcome confounding. Also, our model allows for the existence of unmeasured post-treatment mediator-outcome confounding, corresponding to the rightmost plot of Figure \ref{fig:CausalDiagram}. The key feature of these causal diagrams is that there are no (even observed) confounding factors between $Z$ and $(A,M,Y)$  due to the randomization of $Z$, and $Z$ has no direct causal paths to $M$ and $Y$ because of Assumption \ref{assp:exc-restr}.

\subsection{Parameters of Primary Interest}
\label{subsec:setup-para}
The causal effect of the treatment $A$ on the outcome $Y$ regarding the mediator $M$ can be attributed to two components, 
\[
\delta(a)=Y(a,M(1))-Y(a,M(0)) \;\;\; {\rm and} \;\;\; \xi(a)=Y(1,M(a))-Y(0,M(a)).
\]
The first component $\delta(a)$ quantifies the effect of $A$ on $Y$ through $M$ by hypothetically fixing the treatment to $A=a$ but reversing the mediator from $M(0)$ to $M(1)$. The second component $\xi(a)$ quantifies the direct treatment effect of $A$ on $Y$ when the mediator is fixed at the value it would have been under $A=a$. As stated in \citet{sjolander2009bounds}, $\xi(a)$ answers the question: ``\textit{If the influence of smoking on lipid levels is eliminated, what is the additional benefit in preventing smoking?}'' Conversely, $\delta(a)$ answers the question: ``{\em What is the benefit of preventing smoking that is specifically attributable to its effect on lipid levels?}'' 

We focus on estimation of two population quantities: the average causal mediation effect (ACME; Imai et al., \citeyear{imai2010identification}), also called the natural indirect effect \citep{pearl2001direct}, given by $\bar{\delta}(a)=\E[\delta(a)]$, and the natural direct effect (NDE), defined as $\bar{\xi}(a)=\E[\xi(a)]$. The average treatment effect (ATE) can be decomposed as 
\[
\tau=\E[Y(1,M(1))-Y(0,M(0))] = \bar{\delta}(0)+\bar{\xi}(1)=\bar{\delta}(1)+\bar{\xi}(0).
\]

\subsection{Identification}
\label{subsec:setup-identif}

Assumptions \ref{assp:exc-restr}--\ref{assp:rand} do not guarantee identification of $\bar{\delta}(a)$ and $\bar{\xi}(a)$ due to noncompliance and latent outcome-mediator confounding. Identification of the two parameters of primary interest requires additional strong and untestable ignorability assumptions that have been extensively studied in existing literature; see, for instance,  \citet{pearl2001direct,robins2003semantics,imai2010general,imai2010identification,hafeman2011alternative}, among others. With the help of instrumental variables, it is possible to identify a set of local parameters defined on the subpopulation of the compliers under weaker assumptions \citep{rudolph2021complier}. This is further discussed in Section \ref{sec:bounds-local}. 

Rather than relying on strong identification assumptions, \citet{sjolander2009bounds} addressed partial identification and derived nonparametric bounds of the ACME and the NDE under randomization of treatment assignments using the Balke-Pearl algorithm, which is based on linear programming formulation of the causal problem with binary factors \citep{balke1994counterfactual,balke1995probabilistic}. These bounds do not rely on the ignorability assumptions and can be used to evaluate how sensitive a point estimate is when the identification assumptions are violated \citep{imai2010identification}. Nonetheless, Sj\"{o}lander's bounds ignore potential noncompliance and cannot be applied to our setting. We extend Sj\"{o}lander's results and establish nonparametric bounds for $\bar{\delta}(a)$ and $\bar{\xi}(a)$ under the IV setting in Section \ref{sec:bounds-pop}. Nonparametric bounds on the local causal parameters in the IV setting without defiers are provided in Section \ref{sec:bounds-local}.

\section{Bounds on Population Parameters}
\label{sec:bounds-pop}

\subsection{Bounds under Minimal Assumptions}
Let $p_{yma.z}=\pr(Y=y, M=m, A=a\mid Z=z)$. Note that $p_{yma.z}$ can be consistently estimated from the observed data. The upper and lower bounds of $\bar{\delta}(a)$ and $\bar{\xi}(a)$ under Assumptions \ref{assp:exc-restr}--\ref{assp:rand} are established as follows, where $\max V$ and $\min V$ for a vector $V$ refer to the maximum and the minimum of all the components in $V$. Proofs of the results are given in the Appendix. 

    \begin{align*}
        &\bar{\delta}(1)\geq  \max\left\{\begin{array}{cc}
           -p_{000.1} - p_{001.0} - p_{010.0} - p_{011.1} - p_{100.1} - p_{110.0} \\
            -p_{000.0} - p_{001.1} - p_{010.1} - p_{011.0} - p_{100.0} - p_{110.1}\\
            p_{101.1} + p_{111.1} - 1 
        \end{array}\right\},
        \end{align*}
        
         \begin{align*}
        &\bar{\delta}(1)\leq  \min\left\{\begin{array}{cc}
           p_{000.0} + p_{010.1} + p_{100.0} + p_{101.1} + p_{110.1} + p_{111.0} \\
           p_{000.1} + p_{010.0} + p_{100.1} + p_{101.0} + p_{110.0} + p_{111.1}\\
           -p_{001.1} - p_{011.1} + 1
        \end{array}\right\},
        \end{align*}
        
         \begin{align*}
        &\bar{\delta}(0)\geq  \max\left\{\begin{array}{cc}
           -p_{001.0} - p_{011.1} - p_{100.1} - p_{101.0} - p_{110.0} + p_{111.0} - p_{111.1} \\
           -p_{001.1} - p_{011.0} - p_{100.0} - p_{101.1} - p_{110.1}\\
           p_{000.0} + p_{010.0} + p_{111.0} - 1 
        \end{array}\right\},
        \end{align*}
        
         \begin{align*}
        &\bar{\delta}(0)\leq  \min\left\{\begin{array}{cc}
           p_{000.1} + p_{001.0} + p_{010.0} - p_{011.0} + p_{011.1} + p_{101.0} + p_{111.1} \\
           p_{000.0} + p_{001.1} + p_{010.1} + p_{101.1} + p_{111.0} \\
           -p_{011.0} - p_{100.0} - p_{110.0} + 1
        \end{array}\right\}.
    \end{align*}
    
    \begin{align*}
        &\bar{\xi}(1)\geq  \max\left\{\begin{array}{cc}
            -p_{000.0} - p_{001.1} - p_{010.1} - p_{100.0} - p_{110.0} - p_{111.0} + p_{111.1}\\
            p_{000.0} + p_{111.0} + 2p_{011.0} + p_{101.1} - p_{000.1} - p_{011.1} - 1 \\
            p_{011.0} + p_{101.1} + p_{111.1} - 1
        \end{array}\right\},
        \end{align*}
        
        \begin{align*}
        &\bar{\xi}(1)\leq  \min\left\{\begin{array}{cc}
           p_{000.0} + p_{100.0} + p_{010.0} + p_{011.0} - p_{011.1} + p_{101.1}  + p_{110.1}  \\
           -p_{001.1} - p_{011.0} - p_{100.0} + p_{100.1} - 2p_{111.0} + p_{111.1} + 1\\
           -p_{001.1} - p_{011.1} - p_{111.0} + 1
        \end{array}\right\},
        \end{align*}
        
        \begin{align*}
        &\bar{\xi}(0)\geq  \max\left\{\begin{array}{cc}
           p_{000.1} + p_{100.1} + p_{111.1} + p_{010.0}  - p_{100.0} - p_{111.0} - 1 \\
           p_{000.0} + p_{010.1} - p_{101.0} + p_{101.1} - p_{110.0} + p_{110.1} - 1\\
           p_{000.0} + p_{010.0} - 1 
        \end{array}\right\},
        \end{align*}
        
        \begin{align*}
        &\bar{\xi}(0)\leq  \min\left\{\begin{array}{cc}
           p_{000.0} - p_{000.1} + p_{011.0} - p_{011.1} - p_{100.1} - p_{110.0} + 1 \\
           p_{001.0} - p_{001.1} + p_{010.0} - p_{010.1} - p_{100.0} - p_{110.1} + 1 \\
           -p_{100.0} - p_{110.0} + 1
        \end{array}\right\}.
    \end{align*}
    
The above bounds are derived using the Balke-Pearl algorithm \citep{balke1994counterfactual,balke1995probabilistic}. The upper and lower bounds are in a general form of taking the maximum/minimum value from a set of observed probability entries. Our proofs show that these symbolic entries correspond to the vertices of a (symbolic) polyhedron; see the Appendix for further detail.

It should be noted that the above bounds are not sharp. The tightest bounds of $\bar{\delta}(a)$ and $\bar{\xi}(a)$ under Assumptions \ref{assp:exc-restr}--\ref{assp:rand} involve over $30$ entries and we decide not to present those lengthy expressions in this paper. In fact, the above bounds are the tightest bounds of $\bar{\delta}(a)$ and $\bar{\xi}(a)$ under Assumptions \ref{assp:exc-restr}--\ref{assp:monotone-A}, impliying that for bounding $\bar{\delta}(a)$ and $\bar{\xi}(a)$, no information is gained by assuming treatment-assignment monotonicity. This is because among the entries of the bounds under Assumptions \ref{assp:exc-restr}--\ref{assp:rand}, only those of the bounds under Assumptions \ref{assp:exc-restr}--\ref{assp:monotone-A} are active when Assumption \ref{assp:monotone-A} is true. A similar phenomenon has been observed when bounding the ATE in the IV setting \citep{balke1997bounds,gabriel2023nonparametric}.

\subsection{Bounds under Monotonicity Assumptions}
\label{subsec:setup-addass}

Following \citet{robins1992identifiability} and \citet{sjolander2009bounds}, we consider a few additional monotonicity assumptions, under which the bounds of the ACME and the NDE become more informative and useful when the corresponding assumptions hold.

\begin{assumption}[Treatment-assignment monotonicity]\label{assp:monotone-A}
   $A(1)\geq A(0)$ almost surely.
\end{assumption}

\begin{assumption}[Mediator-treatment monotonicity]\label{assp:monotone-M}
    $M(1)\geq M(0)$ almost surely.
\end{assumption}

\begin{assumption}[Outcome-mediator monotonicity]\label{assp:monotone-Y-a}
    $Y(a,1)\geq Y(a,0)$ almost surely for $a=0,1$.
\end{assumption}

\begin{assumption}[Outcome-treatment monotonicity]\label{assp:monotone-Y-m}
    $Y(1,m)\geq Y(0,m)$ almost surely for $m=0,1$.
\end{assumption}

Assumptions \ref{assp:monotone-A}--\ref{assp:monotone-Y-m} are imposed on the unit level rather than on population averages. They are strong and untestable but may still be reasonable in many real-world applications. For instance, in the smoking example described earlier, imposing Assumption \ref{assp:monotone-Y-a} is plausible if the researcher is confident that, conditional on smoking status, the causal effect of hyperlipidemia on CVD is positive for every individual.

The upper and lower bounds for the ACME and the NDE under different combinations of the additional monotonicity assumptions are presented in the Appendix. It is noted that the upper bounds of $\bar{\delta}(a)$ under Assumptions \ref{assp:exc-restr}--\ref{assp:monotone-M}, \ref{assp:monotone-Y-m} are the same as the ones under Assumptions \ref{assp:exc-restr}--\ref{assp:monotone-Y-m}; see the Appendix for further detail.

\section{Bounds on Local Parameters}
\label{sec:bounds-local}

The IV setting without defiers as specified by Assumptions \ref{assp:exc-restr}--\ref{assp:monotone-A} allows us to identify the local ATE defined on the subpopulation of compliers \citep{angrist1996identification}. More formally, let $\tau^l=\E[Y(1,M(1))-Y(0,M(0))\mid A(1)>A(0)]$. We have 
\begin{align*}
    \tau^l = \frac{\E[Y(1,A(1),M(A(1)))-Y(0,A(0),M(A(0)))]}{\E[A(1)-A(0)]}
    = \frac{\pr(Y=1|Z=1)-\pr(Y=1|Z=0)}{\pr(A=1|Z=1)-\pr(A=1|Z=0)},
\end{align*}
which is exactly the ratio of the average assignment effect of $Z$ on $Y$ to the average assignment effect of $Z$ on $A$. We can similarly define the local average causal mediation effect (LACME) as $\bar{\delta}^l(a)=\E[\delta(a)\mid A(1)>A(0)]$ and the local natural indirect effect (LNDE) as $\bar{\xi}^l(a)=\E[\xi(a)\mid A(1)>A(0)]$. The two local parameters are also nonidentifiable due to latent mediator-outcome confounding. In the following, we derive their tightest upper and lower bounds under the IV setting without defiers. Specifically, we focus on the bounds of $\bar{\delta}^l(a)$, which further lead to the bounds of $\bar{\xi}^l(a)$ by noting that $\tau^l=\bar{\delta}^l(1-a)+\bar{\xi}^l(a)$. 

Let $p_{ym.z}=\pr(Y=y,M=m\mid Z=z, A(1)=1,A(0)=0)$, which can be identified by 
\[
p_{ym.z}=\frac{\pr(Y=y,M=m\mid Z=z)}{\pr(A=1|Z=1)-\pr(A=1|Z=0)}.
\]
Under Assumptions \ref{assp:exc-restr}--\ref{assp:monotone-A}, the sharp bounds of $\bar{\delta}^l(a)$ are given by
        \begin{align*}
        \max\left\{\begin{array}{cc}
           -p_{01.0} - p_{01.1}  - p_{11.0} \\
           -p_{00.0} - p_{00.1} - p_{10.0} \\
           - p_{00.1} - p_{01.1}
           \end{array}\right\}
        \leq\bar{\delta}^l(1)\leq \min\left\{\begin{array}{cc}
           p_{00.0} + p_{10.0} + p_{10.1} \\
           p_{01.0} + p_{11.0} + p_{11.1}\\
           p_{10.1} + p_{11.1}
        \end{array}\right\},
        \end{align*}
    \begin{align*}
    \max\left\{\begin{array}{cc}
           - p_{01.1} - p_{11.0} - p_{11.1} \\
            - p_{00.1} - p_{10.0} - p_{10.1}\\
           -p_{10.0} - p_{11.0} 
        \end{array}\right\}
        \leq\bar{\delta}^l(0)\leq \min\left\{\begin{array}{cc}
           p_{00.0} + p_{00.1} + p_{10.1} \\
           p_{01.0} + p_{01.1} + p_{11.1} \\
           p_{00.0} + p_{01.0}
        \end{array}\right\}.
    \end{align*}

The bounds on the ACME in the setting without noncompliance have an equivalent form as the above bounds \citep{sjolander2009bounds,imai2010identification}, except that $p_{ym.z}=\pr(Y=y,M=m\mid Z=z)$. Therefore, the bounds derived by \citet{sjolander2009bounds} can be actually extended to the IV setting without defiers by simply including the divisor $\pr(A=1|Z=1)-\pr(A=1|Z=0)$. However, when considering bounds under additional monotonicity assumptions, \citet{sjolander2009bounds} assumes that Assumptions \ref{assp:monotone-M}--\ref{assp:monotone-Y-m} are all true. In the Appendix, we derive the bounds of $\bar{\delta}^l(a)$ under different combinations of Assumptions \ref{assp:monotone-M}--\ref{assp:monotone-Y-m}, which essentially complement and generalize the results of \citet{sjolander2009bounds}.

\section{Analysis of the JOBS~\rom{2} Data}
\label{sec:real-data}

We apply the derived bounds to the Job Search Intervention Study (JOBS~\rom{2}) dataset available from the R package \texttt{mediation} \citep{tingley2014mediation}. The JOBS~\rom{2} is a randomized controlled trial designed to evaluate the effect of job training on unemployed participants. The dataset we use here consists of $899$ observations and is part of the original JOBS~\rom{2} data \citep{vinokur1997mastery}.
Among these units, $600$ job seekers were randomly intervened and invited to attend a series of workshops designed to help the participants enhance job-search skills and managing setbacks during job-hunting ($Z=1$), while the remaining individuals only received a booklet containing relatively brief information on job-search methods ($Z=0$). Among the intervened individuals, $228$ dropped out of the workshops ($(A,Z)=(0,1)$), leaving $372$ participants who attended at least one session ($A=1$). Job-search self-efficacy was then assessed through follow-ups: $555$ individuals were classified as having high self-efficacy ($M=1$), and the rest as having low self-efficacy ($M=0$). Whether the job seekers were employed was also recorded, with $Y=1$ indicating employment. Our primary interest in this application is to evaluate the effect of the job workshops on employment mediated through job-search self-efficacy, i.e., the ACME and the NDE.

\begin{table}[h]
  \centering
  \caption{Bounds on the population parameters under different assumptions.}
    \begin{tabular}{lcccc}
    \toprule
          & \multicolumn{4}{c}{Assumptions} \\
\cmidrule{2-5} & \ref{assp:exc-restr}--\ref{assp:rand}  & \ref{assp:exc-restr}--\ref{assp:monotone-A} & \ref{assp:exc-restr}--\ref{assp:monotone-M}  & \ref{assp:exc-restr}--\ref{assp:monotone-A}, \ref{assp:monotone-Y-a} \\
    \midrule
    $\bar{\delta}(1)$ & $[-0.795, 0.585]$ & $[-0.795, 0.585]$ & $[-0.228, 0.228]$ & $[-0.373, 0.292]$  \\
    $\bar{\delta}(0)$ & $[-0.288, 0.678]$ & $[-0.288, 0.678]$ & $[-0.107, 0.175]$ & $[-0.181, 0.328]$ \\
    $\bar{\xi}(1)$ & $[-0.760, 0.585]$ & $[-0.760, 0.585]$ & $[-0.257, 0.404]$ & $[-0.410, 0.478]$  \\
    $\bar{\xi}(0)$ & $[-0.288, 0.712]$ & $[-0.288, 0.712]$ & $[-0.161, 0.376]$ & $[-0.224, 0.441]$  \\
    \bottomrule
    \end{tabular}%
  \label{tab:pop-bounds1}%
\end{table}%

\begin{table}[h]
  \centering
  \caption{Bounds on the population parameters under different assumptions.}
    \begin{tabular}{lcccc}
    \toprule
          & \multicolumn{4}{c}{Assumptions} \\
\cmidrule{2-5} & \ref{assp:exc-restr}--\ref{assp:monotone-A}, \ref{assp:monotone-Y-m}  & \ref{assp:exc-restr}--\ref{assp:monotone-Y-a} & \ref{assp:exc-restr}--\ref{assp:monotone-M}, \ref{assp:monotone-Y-m}  & \ref{assp:exc-restr}--\ref{assp:monotone-Y-m} \\
    \midrule
    $\bar{\delta}(1)$ & $[-0.800, 0.297]$ & $[0.0, 0.228]$ & $[-0.228, 0.144]$ & $[0.0, 0.144]$  \\
    $\bar{\delta}(0)$ & $[-0.288, 0.297]$ & $[0.0, 0.175]$ & $[-0.107, 0.144]$ & $[0.0, 0.144]$ \\
    $\bar{\xi}(1)$ & $[0.0, 0.585]$ & $[-0.257, 0.297]$ & $[0.001, 0.404]$ & $[0.001, 0.297]$  \\
    $\bar{\xi}(0)$ & $[0.0, 0.712]$ & $[-0.161, 0.297]$ & $[0.001, 0.376]$ & $[0.001, 0.297]$  \\
    \bottomrule
    \end{tabular}%
  \label{tab:pop-bounds2}%
\end{table}%

It is apparent that Assumption \ref{assp:monotone-A} is guaranteed by design for this example. Assumption \ref{assp:monotone-Y-a} also seems plausible as improving job-research self-efficacy should not harm employment. The bounds on $\bar{\delta}(a)$ and $\bar{\xi}(a)$ under different assumptions are presented in Tables \ref{tab:pop-bounds1} and \ref{tab:pop-bounds2}. Since Assumption \ref{assp:monotone-A} holds, the bounds for all the causal parameters under Assumptions \ref{assp:exc-restr}--\ref{assp:rand} are the same as the ones under Assumptions \ref{assp:exc-restr}--\ref{assp:monotone-A}, confirming our arguments in Section \ref{sec:bounds-pop}. Note that the computed bounds are generally quite wide, implying that the point estimates are sensitive to violations of the identification assumptions. The computed bounds get tighter when we impose more monotonicity assumptions. When the researchers are confident of Assumptions \ref{assp:exc-restr}--\ref{assp:monotone-M} and \ref{assp:monotone-Y-m}, $\bar{\xi}(a)$ are strictly positive and greater than $0.001$. On the other hand, if the researchers are confident of Assumptions \ref{assp:exc-restr}--\ref{assp:monotone-Y-a}, then $\bar{\delta}(a)$ are non-negative. The bounds on $\bar{\delta}^l(a)$ and $\bar{\xi}^l(a)$ are presented in Tables \ref{tab:local-bounds1} and \ref{tab:local-bounds2}, which are generally more informative compared to the bounds for the population parameters. Particularly, when Assumptions \ref{assp:exc-restr}--\ref{assp:monotone-Y-m} are true, the bounds for the local parameters are very tight and can be used to infer the sign of the parameters.

\begin{table}[h]
  \centering
  \caption{Bounds on the local parameters under different assumptions.}
    \begin{tabular}{lccc}
    \toprule
          & \multicolumn{3}{c}{Assumptions} \\
\cmidrule{2-4} & \ref{assp:exc-restr}--\ref{assp:monotone-A} & \ref{assp:exc-restr}--\ref{assp:monotone-M}  & \ref{assp:exc-restr}--\ref{assp:monotone-A}, \ref{assp:monotone-Y-a} \\
    \midrule
    $\bar{\delta}^l(1)$ & $[-0.669, 0.331]$ & $[-0.126, 0.126]$ & $[-0.231, 0.228]$  \\
    $\bar{\delta}^l(0)$ & $[-0.238, 0.706]$ & $[-0.087, 0.126]$ & $[-0.152, 0.373]$ \\
    $\bar{\xi}^l(1)$ & $[-0.614, 0.331]$ & $[-0.033, 0.179]$ & $[-0.280,  0.244]$  \\
    $\bar{\xi}^l(0)$ & $[-0.238, 0.762]$ & $[-0.033, 0.219]$ & $[-0.136, 0.324]$  \\
    \bottomrule
    \end{tabular}%
  \label{tab:local-bounds1}%
\end{table}%

\begin{table}[h]
  \centering
  \caption{Bounds on the local parameters under different assumptions.}
    \begin{tabular}{lcccc}
    \toprule
          & \multicolumn{4}{c}{Assumptions} \\
\cmidrule{2-5} & \ref{assp:exc-restr}--\ref{assp:monotone-A}, \ref{assp:monotone-Y-m}  & \ref{assp:exc-restr}--\ref{assp:monotone-Y-a} & \ref{assp:exc-restr}--\ref{assp:monotone-M}, \ref{assp:monotone-Y-m}  & \ref{assp:exc-restr}--\ref{assp:monotone-Y-m} \\
    \midrule
    $\bar{\delta}^l(1)$ & $[-0.669,  0.093]$ & $[0.0, 0.126]$ & $[-0.126,  0.077]$ & $[0.0,   0.077]$  \\
    $\bar{\delta}^l(0)$ & $[-0.238,  0.093]$ & $[0.0, 0.126]$ & $[-0.087, 0.077]$ & $[0.0,   0.077]$ \\
    $\bar{\xi}^l(1)$ & $[0.0,    0.331]$ & $[-0.033, 0.093]$ & $[0.016, 0.179]$ & $[0.016, 0.093]$  \\
    $\bar{\xi}^l(0)$ & $[0.0,   0.762]$ & $[-0.033, 0.093]$ & $[0.016, 0.219]$ & $[0.016, 0.093]$  \\
    \bottomrule
    \end{tabular}%
  \label{tab:local-bounds2}%
\end{table}%

\section{Additional Remarks}
\label{sec:remarks}

In this paper, we established nonparametric upper and lower bounds of the ACME and the NDE in the IV setting and extended the results to the LACME and the LNDE in the IV setting without defiers. These bounds provide partial information of the parameters of interest under minimal assumptions. However, the derived bounds are sometimes not very informative for inference on whether the average direct or indirect causal effects are positive or negative, since in many cases the intervals formed by the upper and lower bounds contain zero. Nevertheless, the intervals can be used to assess how sensitive point estimates are to violations of identification assumptions \citep{richardson2015nonparametric}. 

In the presence of pre-treatment covariates, the derived bounds can be further tightened. The key idea is to partition the covariate space into disjoint subspaces so that the bounds within each subspace are as tight as possible. By taking weighted averages of these partition-specific bounds, one can obtain overall bounds that are generally tighter than those derived without incorporating covariate information. Owing to randomization, the averaged bounds remain valid for the parameters of interest. For further details, readers are referred to \citet{liang2025estimation} and \citet{levis2025covariate}.

\medskip

\noindent{\bf Funding information:} This research was supported in part by a grant from the Natural Sciences and Engineering Research Council of Canada.

\bibliography{main.bib}
\bibliographystyle{asa}

\newpage

\appendix

\section*{Appendix}

\subsection*{A \ Derivation of the Bounds}
\label{sec:app-proof}

The bounds presented in the main sections of the paper as well as additional ones to be included in this Appendix are derived based on the Balke-Pearl algorithm outlined in \citet{balke1994counterfactual} and \citet{balke1995probabilistic}. We first briefly review the method. Let $O_y=(Y(1,1),Y(1,0),Y(0,1),Y(0,0))$, $O_m=(M(1),M(0))$, and $O_a=(A(1),A(0))$. The values of $O_{y}$ and $O_m$ are encoded as given respectively in Table \ref{tab:ecode-oyi} and Table \ref{tab:ecode-omi}. Define $q_{ijk}=\pr(O_y=o_{yi},O_m=o_{mj},O_a=o_{mk})$ for $i=0,\cdots,15$, $j,k=0,1,2,3$ (See Table \ref{tab:ecode-oyi} and Table \ref{tab:ecode-omi} for the respective definitions of $o_{yi}$ and $o_{mj}$). 

\begin{table}[htbp]
  \centering
  \caption{The values of $o_{yi}$ for $i=0,\cdots,15$.}
    \begin{tabular}{lrrrrrrrrrrrrrrrr}
    \toprule
    $i$     & 0     & 1     & 2     & 3     & 4     & 5     & 6     & 7     & 8     & 9     & 10    & 11    & 12    & 13    & 14    & 15 \\
    \midrule
    \multirow{4}[1]{*}{$o_{yi}$} & 1     & 1     & 1     & 1     & 1     & 1     & 1     & 1     & 0     & 0     & 0     & 0     & 0     & 0     & 0     & 0 \\
          & 1     & 1     & 1     & 1     & 0     & 0     & 0     & 0     & 1     & 1     & 1     & 1     & 0     & 0     & 0     & 0 \\
          & 1     & 1     & 0     & 0     & 1     & 1     & 0     & 0     & 1     & 1     & 0     & 0     & 1     & 1     & 0     & 0 \\
          & 1     & 0     & 1     & 0     & 1     & 0     & 1     & 0     & 1     & 0     & 1     & 0     & 1     & 0     & 1     & 0 \\
          \bottomrule
    \end{tabular}%
  \label{tab:ecode-oyi}%
\end{table}%

\begin{table}[htbp]
  \centering
  \caption{The values of $o_{mi}$ for $i=0,1,2,3$.}
    \begin{tabular}{lrrrr}
    \toprule
    $i$     & 0     & 1     & 2     & 3 \\
    \midrule
    \multirow{2}[1]{*}{$o_{mi}$} & 1     & 1     & 0     & 0 \\
          & 1     & 0     & 1     & 0 \\
          \bottomrule
    \end{tabular}%
  \label{tab:ecode-omi}%
\end{table}%

Under Assumptions \ref{assp:exc-restr} and \ref{assp:relev}, the parameters of interest are linear in $q_{ijk}$. For example, it follows from simple algebraic manipulations that
\begin{align*}
    \bar\delta(1)&=\E[Y(1,M(1))-Y(1,M(0))]\\
    &= \pr(Y(1,1)=1,M(1)=1) + \pr(Y(1,0)=1,M(1)=0) \\
    & \;\;\;\; - \pr(Y(1,1)=1,M(0)=1) - \pr(Y(1,0)=1,M(0)=0)\\
    & = \sum_{i=4,5,6,7}\sum_{j=1}\sum_{k=0,1,2,3}q_{ijk} + \sum_{i=8,9,10,11}\sum_{j=2}\sum_{k=0,1,2,3}q_{ijk}\\
    &\;\;\;\; - \sum_{i=4,5,6,7}\sum_{j=2}\sum_{k=0,1,2,3}q_{ijk} - \sum_{i=8,9,10,11}\sum_{j=1}\sum_{k=0,1,2,3}q_{ijk}.
\end{align*}
It follows from $\sum_{ijk}q_{ijk}=1$ that the degree of freedom for $q_{ijk}$ is $255$. The density of the observed data $(Y,M,A,Z)$ is given by $\pr(Y=y,M=m,A=a,Z=z) = p(y,m,a,z)=p(y,m,a|z)p(z)$ which is fully characterized by $p(y,m,a|z)=\pr(Y=y,M=m,A=a\mid Z=z)$ and $p(z)=\pr(Z=z)$ for $y,m,a,z\in\{0,1\}$. Denote $p_{yma.z}=p(y,m,a|z)$ as in the main sections of the paper. Under Assumption \ref{assp:rand}, we can show that $p_{yma.z} = \pr(Y(a,m)=y, M(a)=m, A(z)=a)$ impose equality constraints on and further reduce the degree of freedom of $q_{ijk}$. For instance, one can easily show that 
\[
p_{111.1}=\sum_{i=0,1,\cdots,7}\sum_{j=0,1}\sum_{k=0,1}q_{ijk}.
\]

Note that $p(z)$ do not impose any constraints on $q_{ijk}$. Under Assumptions \ref{assp:exc-restr}--\ref{assp:rand}, there are $2^4+1=17$ equality constraints on $q_{ijk}$. The monotonicity assumptions introduce more equality constraints on $q_{ijk}$. For example, Assumption \ref{assp:monotone-A} which states that $A(1)\geq A(0)$ enforces $q_{ijk}=0$ when $k=2$; Assumption \ref{assp:monotone-Y-a} which states that $Y(a,1)\geq Y(a,0)$ enforces $q_{ijk}=0$ when $i=2,6,8,9,10,11,14$. Let $q=(q_{ijk})$ be a $256$-dimensional vector of $q_{ijk}$. For example, computing the lower bound of $\bar\delta(1)$ under Assumptions \ref{assp:exc-restr}--\ref{assp:rand} and possibly others corresponds to solving the symbolic linear programming problem 
\begin{equation}\label{eq:symbolic-lp}
    \begin{aligned}
        &\min_{{q}} c^\top {q}\\
        \text{s.t. }& A{q}=b,\\
        & {q}\geq 0,
    \end{aligned}
\end{equation}
where $\bar{\delta}(1) = c^\top {q}$ and $A{q}=b$ are the equality constraints for $q_{ijk}$. To solve (\ref{eq:symbolic-lp}), we consider its dual problem
\begin{equation}\label{eq:symbolic-dual}
    \begin{aligned}
        &\max_{{w}} b^\top {w}\\
        \text{s.t. }& A^\top{w}\leq c.
    \end{aligned}
\end{equation}
The duality gap is zero because the constraints in (\ref{eq:symbolic-lp}) are all affine (linear). Note that the $P=\{w\mid A^\top{w}\leq c\}$ is a polyhedron. The maximum value of (\ref{eq:symbolic-dual}) is attained at either the vertices or the rays of the polyhedron $P$. Due to the constraints ${q}\geq 0$ and $1^\top{q}=1$, the solution to (\ref{eq:symbolic-lp}) must be bounded and thus the problem (\ref{eq:symbolic-dual}) must also be bounded. Because of the boundedness, there is no need to consider the rays of $P$ and the solution to (\ref{eq:symbolic-dual}) will be 
\begin{align}\label{eq:extre-vertices}
    \max_{w\in V} b^\top w
\end{align}
where $V$ is a collection of the vertices of the polyhedron $P$. To solve (\ref{eq:symbolic-dual}), therefore, we can enumerate all the vertices of the polyhedron $P$ and then substitute into (\ref{eq:extre-vertices}). For the extreme vertex enumeration algorithms, we refer readers to \citet{mattheiss1973algorithm} and \citet{matheiss1980survey}.

\subsection*{B \ Bounds on Population Parameters Under Monotonicity}

We presented the bounds on $\bar{\delta}(a)$ and $\bar{\xi}(a)$ in Section \ref{sec:bounds-pop} under Assumptions \ref{assp:exc-restr}--\ref{assp:rand},  which are also the sharp bounds of $\bar{\delta}(a)$ and $\bar{\xi}(a)$ under Assumptions \ref{assp:exc-restr}--\ref{assp:monotone-A}. With additional monotonicity assumptions given in Section \ref{subsec:setup-addass}, these bounds can be tightened. The results are given below, with Section {\bf B.1} presenting the sharp bounds for $\bar{\delta}(a)$ and Section {\bf B.2} presenting the sharp bounds for $\bar{\xi}(a)$. Proofs of these bounds follow the  Pearl-Balke method as outlined in Section {\bf A} of the Appendix.

\subsubsection*{B.1 \ Bounds on the ACME}

The sharp bounds for $\bar{\delta}(a)$ under different assumptions are established as follows. Under Assumptions \ref{assp:exc-restr}--\ref{assp:monotone-M}, we have
    \begin{align*}
        &\bar{\delta}(1)\geq  \max\left\{\begin{array}{cc}
           p_{001.1} + p_{010.0} + p_{101.1} + p_{110.0} + p_{111.0} - 1 \\
           -p_{000.1} - p_{011.1} - p_{100.1} 
        \end{array}\right\},
        \end{align*}
        \begin{align*}
        &\bar{\delta}(1)\leq  \min\left\{\begin{array}{cc}
           -p_{001.1} - p_{010.0} - p_{011.0} - p_{101.1} - p_{110.0} + 1 \\
           p_{000.1} + p_{100.1} + p_{111.1} 
        \end{array}\right\},
        \end{align*}
        \begin{align*}
        &\bar{\delta}(0)\geq  \max\left\{\begin{array}{cc}
           p_{000.1} + p_{001.1} + p_{010.0} + p_{101.1} + p_{110.0} + p_{111.0} - 1 \\
           -p_{011.0} - p_{100.0}
        \end{array}\right\},
        \end{align*}
        \begin{align*}
        &\bar{\delta}(0)\leq  \min\left\{\begin{array}{cc}
           -p_{001.1} - p_{010.0} - p_{011.0} - p_{100.1} - p_{101.1} - p_{110.0} + 1\\
           p_{000.0} + p_{111.0} 
        \end{array}\right\}.
    \end{align*}

    Under Assumptions \ref{assp:exc-restr}--\ref{assp:monotone-A}, \ref{assp:monotone-Y-a}, we have
    \begin{align*}
        \max\left\{\begin{array}{cc}
           -p_{001.1} - p_{010.1} - p_{110.1} \\
           -p_{001.0} - p_{010.0} - p_{110.0} 
        \end{array}\right\}\leq
        &\bar{\delta}(1)\leq  \min\left\{\begin{array}{cc}
           p_{000.0} + p_{100.0} + p_{111.0} \\
           p_{000.1} + p_{100.1} + p_{111.1}
        \end{array}\right\},
        \end{align*}
        \begin{align*}
        \max\left\{\begin{array}{cc}
           -p_{001.1} - p_{101.1} - p_{110.1} \\
           -p_{001.0} - p_{101.0} - p_{110.0}
        \end{array}\right\}\leq
        &\bar{\delta}(0)\leq  \min\left\{\begin{array}{cc}
           p_{000.0} + p_{111.0} \\
           p_{000.1} - p_{011.0} + p_{011.1} + p_{111.1}
        \end{array}\right\}.
    \end{align*}

    Under Assumptions \ref{assp:exc-restr}--\ref{assp:monotone-A}, \ref{assp:monotone-Y-m}, we have
    \begin{align*}
        &\bar{\delta}(1)\geq\max\left\{\begin{array}{cc}
           -p_{000.0} - p_{001.1} - p_{010.1} - p_{011.0} - p_{100.0} - p_{110.1} \\
           -p_{000.1} - p_{001.0} - p_{010.0} - p_{011.1} - p_{100.1} - p_{110.0}\\
           p_{101.1} + p_{111.1} - 1 
        \end{array}\right\},
        \end{align*}
        \begin{align*}
        &\bar{\delta}(1)\leq  \min\left\{\begin{array}{cc}
           p_{000.1} + p_{010.0} + p_{101.0} + p_{111.1} \\
           p_{000.0} + p_{010.1} + p_{101.1} + p_{111.0}\\
           -p_{001.1} - p_{011.1} - p_{100.0} - p_{110.0} + 1
        \end{array}\right\},
        \end{align*}
        \begin{align*}
        &\bar{\delta}(0)\geq\max\left\{\begin{array}{cc}
           -p_{001.0} - p_{011.1} - p_{100.1} - p_{101.0} - p_{110.0} + p_{111.0} - p_{111.1} \\
           -p_{001.1} - p_{011.0} - p_{100.0} - p_{101.1} - p_{110.1} \\
           p_{000.0} + p_{010.0} + p_{111.0} - 1
        \end{array}\right\},
        \end{align*}
        \begin{align*}
        &\bar{\delta}(0)\leq  \min\left\{\begin{array}{cc}
           p_{000.0} + p_{010.1} + p_{101.1} + p_{111.0} \\
           p_{000.1} + p_{010.0} + p_{101.0} + p_{111.1}\\
           -p_{001.1} - p_{011.1} - p_{100.0} - p_{110.0} + 1
        \end{array}\right\}.
    \end{align*}

    Under Assumptions \ref{assp:exc-restr}--\ref{assp:monotone-Y-a}, we have
    \begin{align*}
        0\leq
        &\bar{\delta}(1)\leq  \min\left\{\begin{array}{cc}
           -p_{001.1} - p_{010.0} - p_{011.0} - p_{101.1} - p_{110.0} + 1\\
           p_{000.1} + p_{100.1} + p_{111.1}   
        \end{array}\right\},
        \end{align*}
        \begin{align*}
        0\leq
        &\bar{\delta}(0)\leq  \min\left\{\begin{array}{cc}
           -p_{001.1} - p_{010.0} - p_{011.0} - p_{100.1} - p_{101.1} - p_{110.0} + 1 \\
           p_{000.0} + p_{111.0}
        \end{array}\right\}.
    \end{align*}

    Under Assumptions \ref{assp:exc-restr}--\ref{assp:monotone-M}, \ref{assp:monotone-Y-m}, we have
    \begin{align*}
        &\bar{\delta}(1)\geq  \max\left\{\begin{array}{cc}
           p_{001.1} + p_{010.0} + p_{101.1} + p_{110.0} + p_{111.0} - 1 \\
           -p_{000.1} - p_{011.1} - p_{100.1}
        \end{array}\right\},
        \end{align*}
        \begin{align*}
        &\bar{\delta}(1)\leq  \min\left\{\begin{array}{cc}
           -p_{001.1} - p_{010.0} - p_{011.0} - p_{100.1} - p_{101.1} - p_{110.0} + 1 \\
           -p_{001.1} - p_{010.1} - p_{011.1} - p_{100.0} - p_{101.0} - p_{110.0} + 1\\
           p_{000.0} + p_{001.0} - p_{001.1} + p_{111.0}\\
           p_{000.1} - p_{110.0} + p_{110.1} + p_{111.1} 
        \end{array}\right\},
        \end{align*}
        \begin{align*}
        &\bar{\delta}(0)\geq  \max\left\{\begin{array}{cc}
           p_{000.1} + p_{001.1} + p_{010.0} + p_{101.1} + p_{110.0} + p_{111.0} - 1 \\
           -p_{011.0} - p_{100.0}
        \end{array}\right\},
        \end{align*}
        \begin{align*}
        &\bar{\delta}(0)\leq  \min\left\{\begin{array}{cc}
           -p_{001.1} - p_{010.0} - p_{011.0} - p_{100.1} - p_{101.1} - p_{110.0} + 1 \\
           -p_{001.1} - p_{010.1} - p_{011.1} - p_{100.0} - p_{101.0} - p_{110.0} + 1\\
           p_{000.0} + p_{001.0} - p_{001.1} + p_{111.0}\\
           p_{000.1} - p_{110.0} + p_{110.1} + p_{111.1}
        \end{array}\right\}.
    \end{align*}

    Under Assumptions \ref{assp:exc-restr}--\ref{assp:monotone-Y-m}, we have
    \begin{align*}
        0\leq
        &\bar{\delta}(1)\leq  \min\left\{\begin{array}{cc}
           -p_{001.1} - p_{010.0} - p_{011.0} - p_{100.1} - p_{101.1} - p_{110.0} + 1\\
           -p_{001.1} - p_{010.1} - p_{011.1} - p_{100.0} - p_{101.0} - p_{110.0} + 1\\
           p_{000.1} - p_{110.0} + p_{110.1} + p_{111.1}\\
           p_{000.0} + p_{001.0} - p_{001.1} + p_{111.0} 
        \end{array}\right\},
        \end{align*}
        \begin{align*}
        0 \leq
        &\bar{\delta}(0)\leq  \min\left\{\begin{array}{cc}
           -p_{001.1} - p_{010.1} - p_{011.1} - p_{100.0} - p_{101.0} - p_{110.0} + 1 \\
           -p_{001.1} - p_{010.0} - p_{011.0} - p_{100.1} - p_{101.1} - p_{110.0} + 1\\
           p_{000.0} + p_{001.0} - p_{001.1} + p_{111.0}\\
           p_{000.1} - p_{110.0} + p_{110.1} + p_{111.1}
        \end{array}\right\}.
    \end{align*}
    
As we impose more assumptions, the derived bounds are expected to be tighter because the parameter space is reduced. Even if not guaranteed by design,  the monotonicity assumptions are considered mild and plausible in many practical scenarios. If the imposed assumptions do not hold, the derived bounds will likely become invalid and not necessarily be tighter, but they can still be useful as a tool for sensitivity analysis. The process of imposing more and more assumptions is similar to the procedure of changing sensitivity parameters in sensitivity analysis. 

\subsubsection*{B.2 \ Bounds on the NDE}

    The sharp bounds for $\bar{\xi}(a)$ under different assumptions are established as follows. Under Assumptions \ref{assp:exc-restr}--\ref{assp:monotone-M}, we have
    \begin{align*}
        &\bar{\xi}(1)\geq  \max\left\{\begin{array}{cc}
           -p_{001.0} + p_{001.1} + p_{010.0} + p_{011.0} - p_{100.0} + p_{100.1} - p_{101.0} + 2p_{101.1} + p_{111.1} - 1 \\
           p_{010.0} + p_{011.0} + p_{101.1} + p_{111.1} - 1 
        \end{array}\right\},
        \end{align*}
        \begin{align*}
        &\bar{\xi}(1)\leq  \min\left\{\begin{array}{cc}
           p_{000.0} - p_{000.1} + p_{001.0} - 2p_{001.1} - p_{011.1} + p_{101.0} - p_{101.1} - p_{110.0} - p_{111.0} + 1 \\
           -p_{001.1} - p_{011.1} - p_{110.0} - p_{111.0} + 1
        \end{array}\right\},
        \end{align*}
        \begin{align*}
        &\bar{\xi}(0)\geq  \max\left\{\begin{array}{cc}
           p_{000.0} + 2p_{010.0} - p_{010.1} + p_{011.0} - p_{011.1} + p_{101.1} + p_{110.0} - p_{110.1} - 1 \\
           p_{000.0} + p_{010.0} + p_{101.1} - 1 
        \end{array}\right\},
        \end{align*}
        \begin{align*}
        &\bar{\xi}(0)\leq  \min\left\{\begin{array}{cc}
           -p_{001.1} - p_{010.0} + p_{010.1} - p_{100.0} - 2p_{110.0} + p_{110.1} - p_{111.0} + p_{111.1} + 1\\
           -p_{001.1} - p_{100.0} - p_{110.0} + 1
        \end{array}\right\}.
    \end{align*}

    Under Assumptions \ref{assp:exc-restr}--\ref{assp:monotone-A}, \ref{assp:monotone-Y-a}, we have
    \begin{align*}
        &\bar{\xi}(1)\geq\max\left\{\begin{array}{cc}
           -p_{000.1} - p_{001.0} + p_{011.0} - p_{011.1} - p_{100.0} - p_{101.0} + p_{101.1} - p_{110.0} \\
           p_{010.0} + p_{011.0} + p_{101.1} + p_{111.1} - 1
        \end{array}\right\},
        \end{align*}
        \begin{align*}
        &\bar{\xi}(1)\leq  \min\left\{\begin{array}{cc}
           p_{000.0} + p_{010.0} + p_{011.0} - p_{011.1} + p_{101.1} + p_{110.1} \\
           -p_{001.1} - p_{011.1} - p_{100.0} - p_{111.0} + 1
        \end{array}\right\},
        \end{align*}
        \begin{align*}
        &\bar{\xi}(0)\geq\max\left\{\begin{array}{cc}
           -p_{001.1} + p_{010.0} - p_{010.1} + p_{011.0} - p_{011.1} - p_{100.0} - p_{110.1} - p_{111.0} \\
           p_{000.0} + p_{010.0} + p_{011.0} + p_{101.1} - 1
        \end{array}\right\},
        \end{align*}
        \begin{align*}
        &\bar{\xi}(0)\leq  \min\left\{\begin{array}{cc}
           p_{000.1} + p_{001.0} + p_{010.0} - p_{100.0} + p_{100.1} + p_{101.1} + p_{111.1} \\
           -p_{011.1} - p_{100.0} - p_{110.0} + 1
        \end{array}\right\}.
    \end{align*}

    Under Assumptions \ref{assp:exc-restr}--\ref{assp:monotone-A}, \ref{assp:monotone-Y-m}, we have
    \begin{align*}
        &\bar{\xi}(1)\geq\max\left\{\begin{array}{cc}
           -p_{000.0} + p_{000.1} - p_{100.0} + p_{100.1} - p_{110.0} + p_{110.1} - p_{111.0} + p_{111.1} \\
           -p_{010.0} + p_{010.1} - p_{100.0} + p_{100.1} - p_{101.0} + p_{101.1} - p_{110.0} + p_{110.1} \\
           0
        \end{array}\right\},
        \end{align*}
        \begin{align*}
        &\bar{\xi}(1)\leq  \min\left\{\begin{array}{cc}
           -p_{001.0} - p_{011.1} - p_{101.0} + p_{101.1} - p_{110.0} + p_{110.1} - p_{111.0} + 1 \\
           -p_{001.1} - p_{011.0} - p_{100.0} + p_{100.1} - 2p_{111.0} + p_{111.1} + 1 \\
           -p_{001.1} - p_{011.1} - p_{111.0} + 1
        \end{array}\right\},
        \end{align*}
        \begin{align*}
        &\bar{\xi}(0)\geq\max\left\{\begin{array}{cc}
           -p_{010.0} + p_{010.1} - p_{100.0} + p_{100.1} - p_{101.0} + p_{101.1} - p_{110.0} + p_{110.1} \\
           p_{001.0} - p_{001.1} + p_{010.0} - p_{010.1} + p_{011.0} - p_{011.1} + p_{101.0} - p_{101.1} \\
           0
        \end{array}\right\},
        \end{align*}
        \begin{align*}
        &\bar{\xi}(0)\leq  \min\left\{\begin{array}{cc}
           p_{001.0} - p_{001.1} + p_{010.0} - p_{010.1} - p_{100.0} - p_{110.1} + 1 \\
           p_{000.0} - p_{000.1} + p_{011.0} - p_{011.1} - p_{100.1} - p_{110.0} + 1 \\
          -p_{100.0} - p_{110.0} + 1
        \end{array}\right\}.
    \end{align*}

    Under Assumptions \ref{assp:exc-restr}--\ref{assp:monotone-Y-a}, we have
    \begin{align*}
        &\bar{\xi}(1)\geq\max\left\{\begin{array}{cc}
           -p_{001.0} + p_{001.1} + p_{010.0} + p_{011.0} - p_{100.0} + p_{100.1} - p_{101.0} + 2p_{101.1} + p_{111.1} - 1 \\
           p_{010.0} + p_{011.0} + p_{101.1} + p_{111.1} - 1 
        \end{array}\right\},
        \end{align*}
        \begin{align*}
        &\bar{\xi}(1)\leq  -p_{001.1} - p_{011.1} - p_{100.0} - p_{110.0} - p_{111.0} + 1,\\
        &\bar{\xi}(0)\geq\max\left\{\begin{array}{cc}
           p_{000.0} + 2p_{010.0} - p_{010.1} + 2p_{011.0} - p_{011.1} + p_{101.1} + p_{110.0} - p_{110.1} - 1 \\
           p_{000.0} + p_{010.0} + p_{011.0} + p_{101.1} - 1
        \end{array}\right\},
        \end{align*}
        \begin{align*}
        &\bar{\xi}(0)\leq  -p_{001.1} - p_{011.1} - p_{100.0} - p_{110.0} + 1.
    \end{align*}

    Under Assumptions \ref{assp:exc-restr}--\ref{assp:monotone-M}, \ref{assp:monotone-Y-m}, we have
    \begin{align*}
        &\bar{\xi}(1)\geq  \max\left\{\begin{array}{cc}
           p_{010.0} - p_{010.1} + p_{011.0} - p_{011.1} - p_{100.0} + p_{100.1} - p_{101.0} + p_{101.1} \\
           -p_{100.0} + p_{100.1} - p_{101.0} + p_{101.1} \\
           p_{010.0} - p_{010.1} + p_{011.0} - p_{011.1} \\
           0
        \end{array}\right\},
        \end{align*}
        \begin{align*}
        &\bar{\xi}(1)\leq  \min\left\{\begin{array}{cc}
           p_{000.0} - p_{000.1} + p_{001.0} - 2p_{001.1} - p_{011.1} + p_{101.0} - p_{101.1} - p_{110.0} - p_{111.0} + 1 \\
           -p_{001.1} - p_{011.1} - p_{110.0} - p_{111.0} + 1
        \end{array}\right\},
        \end{align*}
        \begin{align*}
        &\bar{\xi}(0)\geq  \max\left\{\begin{array}{cc}
           p_{010.0} - p_{010.1} + p_{011.0} - p_{011.1} - p_{100.0} + p_{100.1} - p_{101.0} + p_{101.1} \\
           -p_{100.0} + p_{100.1} - p_{101.0} + p_{101.1} \\
           p_{010.0} - p_{010.1} + p_{011.0} - p_{011.1} \\
           0
        \end{array}\right\},
        \end{align*}
        \begin{align*}
        &\bar{\xi}(0)\leq  \min\left\{\begin{array}{cc}
           -p_{001.1} - p_{010.0} + p_{010.1} - p_{100.0} - 2p_{110.0} + p_{110.1} - p_{111.0} + p_{111.1} + 1 \\
           -p_{001.1} - p_{100.0} - p_{110.0} + 1
        \end{array}\right\}.
    \end{align*}

    Under Assumptions \ref{assp:exc-restr}--\ref{assp:monotone-Y-m}, we have
    \begin{align*}
        &\bar{\xi}(1)\geq  \max\left\{\begin{array}{cc}
           p_{010.0} - p_{010.1} + p_{011.0} - p_{011.1} - p_{100.0} + p_{100.1} - p_{101.0} + p_{101.1} \\
           -p_{100.0} + p_{100.1} - p_{101.0} + p_{101.1} \\
           p_{010.0} - p_{010.1} + p_{011.0} - p_{011.1} \\
           0
        \end{array}\right\},
        \end{align*}
        \begin{align*}
        &\bar{\xi}(1)\leq  -p_{001.1} - p_{011.1} - p_{100.0} - p_{110.0} - p_{111.0} + 1,
        \end{align*}
        \begin{align*}
        &\bar{\xi}(0)\geq  \max\left\{\begin{array}{cc}
           p_{010.0} - p_{010.1} + p_{011.0} - p_{011.1} - p_{100.0} + p_{100.1} - p_{101.0} + p_{101.1} \\
           -p_{100.0} + p_{100.1} - p_{101.0} + p_{101.1} \\
           p_{010.0} - p_{010.1} + p_{011.0} - p_{011.1} \\
           0
        \end{array}\right\},
        \end{align*}
        \begin{align*}
        &\bar{\xi}(0)\leq  -p_{001.1} - p_{011.1} - p_{100.0} - p_{110.0} + 1.
    \end{align*}

\subsection*{C \ Bounds on Local Parameters Under Monotonicity}

For local causal parameters, we have established sharp bounds on $\bar{\delta}^l(a)$ under Assumptions \ref{assp:exc-restr}--\ref{assp:monotone-A} in Section \ref{sec:bounds-local}. In this section of the Appendix, we present sharp bounds of $\bar{\delta}^l(a)$ under additional monotonicity assumptions.
Under Assumptions \ref{assp:exc-restr}--\ref{assp:monotone-M}, we have
        \begin{align*}
        \max\left\{\begin{array}{cc}
           -p_{00.0} + p_{00.1} - p_{10.0} + p_{10.1}\\
           - p_{01.1}  
        \end{array}\right\}
        \leq\bar{\delta}^*(1)\leq \min\left\{\begin{array}{cc}
           p_{00.0} - p_{00.1} + p_{10.0} - p_{10.1} \\
           p_{11.1}
        \end{array}\right\},
        \end{align*}
        \begin{align*}
        \max\left\{\begin{array}{cc}
           -p_{00.0} + p_{00.1} - p_{10.0} + p_{10.1} \\
           -p_{10.0} 
        \end{array}\right\}
        \leq\bar{\delta}^*(0)\leq \min\left\{\begin{array}{cc}
           p_{00.0}  - p_{00.1} + p_{10.0} - p_{10.1} \\
           p_{00.0}
        \end{array}\right\}.
    \end{align*}

    Under Assumptions \ref{assp:exc-restr}--\ref{assp:monotone-A}, \ref{assp:monotone-Y-a}, we have              
        \begin{align*}
        \max\left\{\begin{array}{cc}
           -p_{01.0} - p_{11.0} \\
           - p_{00.1}
        \end{array}\right\}
        \leq\bar{\delta}^l(1)\leq \min\left\{\begin{array}{cc}
           p_{00.0} + p_{10.0}\\
           p_{11.1}
        \end{array}\right\},
        \end{align*}
        \begin{align*}
        \max\left\{\begin{array}{cc}
           - p_{00.1} - p_{10.1} \\
           -p_{11.0}
        \end{array}\right\}
        \leq\bar{\delta}^l(0)\leq \min\left\{\begin{array}{cc}
           p_{01.1} + p_{11.1} \\
           p_{00.0} 
        \end{array}\right\}.
    \end{align*}

    Under Assumptions \ref{assp:exc-restr}--\ref{assp:monotone-A}, \ref{assp:monotone-Y-m}, we have
        \begin{align*}
        \max\left\{\begin{array}{cc}
           -p_{01.0} - p_{01.1} - p_{11.0}  \\
           -p_{00.0} - p_{00.1} - p_{10.0} \\
           - p_{00.1} - p_{01.1}
        \end{array}\right\}
        \leq\bar{\delta}^l(1)\leq \min\left\{\begin{array}{cc}
           p_{00.0} - p_{00.1} + p_{01.0} - p_{01.1} \\
           p_{00.0} + p_{10.1}\\
           p_{01.0} + p_{11.1} 
        \end{array}\right\},
        \end{align*}
        \begin{align*}
        \max\left\{\begin{array}{cc}
           - p_{01.1} - p_{11.0} - p_{11.1} \\
           - p_{00.1} - p_{10.0} - p_{10.1} \\
           -p_{10.0}  - p_{11.0} 
        \end{array}\right\}
        \leq\bar{\delta}^l(0)\leq \min\left\{\begin{array}{cc}
           p_{00.0} - p_{00.1} + p_{01.0} - p_{01.1} \\
           p_{00.0} + p_{10.1}\\
           p_{01.0} + p_{11.1}
        \end{array}\right\}.
    \end{align*}

    Under Assumptions \ref{assp:exc-restr}--\ref{assp:monotone-Y-a}, we have
    \begin{align*}
        0\leq
        &\bar{\delta}^l(1)\leq \min\left\{\begin{array}{cc}
           p_{00.0} - p_{00.1} + p_{10.0} - p_{10.1}\\
           p_{11.1}
        \end{array}\right\},
        \end{align*}
        \begin{align*}
        0 \leq
        &\bar{\delta}^l(0)\leq \min\left\{\begin{array}{cc}
           p_{00.0} - p_{00.1} + p_{10.0} - p_{10.1} \\
           p_{00.0} 
     \end{array}\right\}.
    \end{align*}

    Under Assumptions \ref{assp:exc-restr}--\ref{assp:monotone-M}, \ref{assp:monotone-Y-m}, we have
        \begin{align*}
        \max\left\{\begin{array}{cc}
           -p_{00.0} + p_{00.1} - p_{10.0} + p_{10.1} \\
           - p_{01.1} 
        \end{array}\right\}
        \leq\bar{\delta}^l(1)\leq \min\left\{\begin{array}{cc}
           p_{00.0} - p_{00.1} + p_{10.0} - p_{10.1} \\
           p_{00.0} - p_{00.1} + p_{01.0} - p_{01.1} \\
           p_{00.0} - p_{00.1} \\
           -p_{11.0} + p_{11.1} 
        \end{array}\right\},
        \end{align*}
        \begin{align*}
        \max\left\{\begin{array}{cc}
           -p_{00.0} + p_{00.1} - p_{10.0} + p_{10.1} \\
           -p_{10.0} 
        \end{array}\right\}
        \leq\bar{\delta}^l(0)\leq \min\left\{\begin{array}{cc}
          p_{00.0} - p_{00.1} + p_{10.0} - p_{10.1} \\
          p_{00.0} - p_{00.1} + p_{01.0} - p_{01.1}\\
          -p_{11.0} + p_{11.1}  \\
          p_{00.0} - p_{00.1}    
        \end{array}\right\}.
    \end{align*}

    Under Assumptions \ref{assp:exc-restr}--\ref{assp:monotone-Y-m}, we have
    \begin{align*}
        0\leq
        &\bar{\delta}^l(1)\leq \min\left\{\begin{array}{cc}
           p_{00.0} - p_{00.1} + p_{01.0} - p_{01.1}\\
           p_{00.0} - p_{00.1} + p_{10.0} - p_{10.1}\\
           p_{00.0} - p_{00.1}\\
           -p_{11.0} + p_{11.1}
        \end{array}\right\},
        \end{align*}
        \begin{align*}
        0\leq
        &\bar{\delta}^l(0)\leq \min\left\{\begin{array}{cc}
           p_{00.0} - p_{00.1} + p_{01.0} - p_{01.1} \\
           p_{00.0} - p_{00.1} + p_{10.0} - p_{10.1}\\
           p_{00.0}  - p_{00.1} \\
           -p_{11.0} + p_{11.1}
        \end{array}\right\}.
    \end{align*}

The above bounds are the tightest bounds under corresponding assumptions. Similarly, they are tighter as we impose more assumptions. It is worth noting that the probability entry $p_{00.0} - p_{00.1} + p_{01.0} - p_{01.1}$ frequently appears in the upper bounds of $\bar{\delta}^l(1)$ and $\bar{\delta}^l(0)$ which is exactly
\[
p_{00.0} - p_{00.1} + p_{01.0} - p_{01.1}=\tau^l.
\]

\end{document}